\documentstyle[12pt]{article}
\def\be{\begin{equation}}
\def\ee{\end{equation}}
\begin{document}

 \title{The Extended Supersymmetrization of the Multicomponent
                        Kadomtsev--Petviashvilli Hierarchy}

                        \author{by \\
                        Ziemowit Popowicz\\ \\
        Institute of Theoretical Physics, University of Wroc{\l}aw\\
        Pl. M. Borna 9 50 - 205 Wroc{\l}aw Poland}
\date{}
\maketitle

\begin{abstract} We describe three different approaches to the
extended (N=2) supersymmetrization of the multicomponent KP
hierarchy.
In the first one we utilize  only superfermions  while in the second
only superbosons and in the third superbosons as well as
superfermions.
It is shown that many soliton equations can be embedded in the
supersymmetry theory by using the first approach even if we do not
change
these
equations in the bosonic limit of the supersymmetry.  In
the second or third approach we obtain a  generalization of
the soliton equations in the bosonic limit which remains
in the class of the
usual commuting functions.  As the byproduct of our analysis we
prove that for the first procedure the bosonic part of
the one--component supersym\-me\-tric KP hierarchy coincides with the
usual classical two--component KP hierarchy.

\end{abstract}

\newpage

\section{Introduction. }
Integrable Hamiltonian systems occupy an important place in diverse
branches of theoretical physics as exactly solvable models of
fundamental
physical phenomena ranging from nonlinear hydrodynamics to string
theory
[1,2,3].
The general Kadomtsev - Petviashvilli (KP) system  [4,5] is  1+1
dimensional integrable model  containing an infinite number of
fields.  In
the Sato approach [6,7,8], the KP hierarchy  is described by the
isospectral
deformations of the  eigenvalue problem  $L\Psi = \lambda\psi$
             for the
pseudodifferential operator
$L = \partial + U_{2} \partial^{-1} + U_{3}\partial^{-2}$
which is given by
\be
L_{t_{n}} = \left[ B_{n}, L\right],
\ee
where $n=2,3,....$  and $B_{n}$   is the differential part of the
microdifferential
operator $L^{n}$.  If we require that $L$ satisfies the
additional condition that
$L^{n} = B_{n},n\geq  2 $
 then the hierarchy of equation given by Eq. (1) are
reduced to the hierarchy of (1+1) dimensional integrable systems
called the
n-reduced KP hierarchy.  For example, the Korteweg - de Vries
equation and the
Boussinesq equation belong to the two-reduced and three-reduced KP
hierarchies, respectively.

        On the other side, a new type of reduction have been proposed
recently
in a series of articles, which leads many (2+1) dimensional
integrable systems
to (1+1) dimensional integrable systems  [9--15].  For example, by
assuming
that L satisfies the constraints
\be
L^{n} = B_{n} + q \partial^{-1} r,
\ee
we can obtain the so called k-constrained KP hierarchy.
Interestingly,
the
one - constrained KP hierarchy coincide with the AKNS hierarchy, and
the
two - constrained KP hierarchy coincides with the Yajima - Oikawa
[16]
hierarchy.  The
k-constrained KP hierarchy was shown to possess Lax pairs, recursion
operators
and the bi-Hamiltonian structures [15].

        However this  classification does not exhaust the known
generalizations of the KP hierarchies. In this paper we consider
two different generalizations of the KP hierarchies. In the
next section we describe the so called multicomponent KP hierarchy
and in
the next chapters we consider the extended supersymmetrization of the
multicomponent KP hierarchies.

        The idea to use the extended supersymmetry (SUSY) for the
generalization of the soliton equations appeared almost in parallel
to
the usage of the SUSY in the quantum field theory [17,18].  The main
idea of SUSY is to treat  bosons and fermions operators equally.  The
first
results, concerned the construction of classical field theories with
fermionic and bosonic fields depending on time and one space variable,
can be found in [19--22].  In many cases, the addition of fermion
fields
does not guarantee that the final theory becomes the SUSY invariant.
Therefore this method was named the fermionic extension in order to
distinguish it from the fully SUSY way.

        We have at the moment many different procedures [23 -48] of
the
supersymmetrization of the soliton equations.  From the soliton point
of
view we can distinguish two different recipes.  In the first we add
to the
theory the new anticommuting Grassmann valued functions only while in
the
second case we also add the new commuting functions.
Interestingly enough it appeared that during the supersymmetrizations,
some typical SUSY effects (compared to the classical theory)
occurred.
We mention few of them ;
the nonuniqueness of the roots for the SUSY Lax operator [36,40],
the lack of the bosonic reduction to the classical equations [35] and
the nonexistence of the extended SUSY extension of the $SL(2,C)$ Kac-
Moody
algebra [30].  These effect strongly relies on the descriptions of
the
generalized classical systems of equations which we would like to
supersymmetrize.  In the classical case the AKNS hierarchy is
connected
with the one--component KP hierarchy and in the SUSY case it is
tempting
to use similar
arguments and try to construct the SUSY version of AKNS hierarchy.
However we are not able to show that the supersymmetric extension
constitutes
the SUSY bihamiltonian system.  Therefore, we do not consider here
the problem of the extended supersymmetrization of the AKNS
hierarchy.
We show that the lack of the existence of the bihamiltonian structure
in our approach is closely connected with the nonexistence of the
extended SUSY $SL(2,C)$ Kac-Moody algebra.

        From this classification of the supersymmetrization methods
one can
infer that the second approach is more important then the first
because we
extend our knowledge on the new commuting functions.  However, it is
not
completely true.  As we show in this paper, it is possible to carry
out the
supersymmetrization of the one--component KP hierarchy in two
different
ways.
We show that despite of using the
superfermions in the
first approach
 for the supersymmetrization of the one--component KP
hierarchy,  the bosonic sector coincides with the usual classical
two--component KP hierarchy.  Interestingly the bosonic part of the
SUSY Lax pair of the one--component KP hierarchy is matrix valued
operator in contrast to the scalar Lax operator in the classical
case.

        The paper is organized as follows.  In the first section we
describe
the multicomponent KP hierarchy.  The second contains the
introduction to the
supersymmetrization of this hierarchy which is  developed in the next
chapters.  More precisely in the third section we describe the
superfermionic
approach while in the fourth the superbosonic.  We use superfermions
as well as
superbosons  in the fifth chapter in the supersymmetrization of our
multicomponent KP hierarchy in order to demonstrate the third (mixed)
possibilities.  The last  section contains  concluding remarks.

        All calculations presented in this paper have been obtained
by the
extensive applications of the symbolic computation language REDUCE.

\section{Multicomponent KP hierarchy. }

        The multicomponent KP hierarchy have been introduced by
Sidorenko
and Strampp [14] which is a straightforward generalization
of the scalar case.
This is a hierarchy associated with the following Lax operator
\be
L_{n} = \partial^{n} + u_{n-2} \partial^{n-2} + \ldots
+ u_{0} + \sum\limits^{m}_{u=1} q_{i}\partial^{-1} r_{i} \, ,
\ee

The corresponding flows could be constructed by means of fractional
power
method [5].  For n=1, one has multicomponent AKNS hierarchy, which
includes
coupled Nonlinear Schr\"{o}dinger [49] equation as an example.
For the case n=2
and n=3 one has the multicomponent Yajima - Oikawa [16] and Melnikov
hierarchy [50] respectively.  We first consider  multicomponent AKNS
hierarchy
which is given by
\be
L = \partial + \sum\limits^{n}_{i=1} q_{i} \partial^{-1}r_{i} \,
,
\ee
and the flows are
\be
L_{t_{k}} = \left[ \left( L^{k}\right)_{+}, L \right] \,  .
\ee
\
\
The bi-hamiltonian structure of these equations have been widely
discussed
in the literature recently [14,51,52] and it has the following
representation
\be
q_{t_{k}} = B^{0} { \delta H_{k+1} \over \delta q}
= B^{1} { \delta H_{k} \over \delta q} \, ,
\ee
where $q=(q_{1}, q_{2},...q_{n}, r_{1},r_{2}, \ldots r_{n})$  and
\be
B^{0} = \pmatrix{O & I \cr -I & O} \, ,
\ee
where $I$   is $m \times m$ identity matrix.
$B^{1}$    is in the form  [52]
\be
B^{1} = \pmatrix{B^{1}_{11}, & B^{1}_{12} \cr\cr
B^{1}_{21}, &B^{1}_{22}}
\, ,
\ee
where    $B^{1}_{n,k} (n,k=1,2)$ are
$m \times m$ matrices with the entries
\be
B^{1}_{11} = \left\{ q_{i} \partial^{-1} q_{j} +
q_{j} \partial^{-1} q_{i} \right\} \, ,
B^{1}_{12} = \left\{ \left( \partial - \sum\limits^{m}_{s=1}
q_{s} \partial^{-1} r_{s}\right) \delta_{ij} -
q_{i} \partial^{-1} r_{j} \right\}
\ee

\be
(B^{1}_{12})^* = - B^{1}_{21} \, ,
\qquad \qquad
B^{1}_{22} = \left\{ r_{i} \partial^{-1} r_{j} + r_{j}
\partial^{-1} r_{i} \right\}
\ee
and $*$ denotes the hermitean conjugation.  In the special case $m=1$
we obtain
\be
B^{1} = \pmatrix{2 q \partial^{-1} q, & \partial - 2 q \partial^{-1}
r \cr\cr
\partial - 2 r \partial^{-1} q, & 2 r \partial^{-1} r}
\, ,
\ee
Interestingly, this Hamiltonian operator could be considered as the
outcome
of the Dirac reduction of the hamiltonian operator connected with the
$SL(2,C)$ Kac-Moody algebra [39].

        The Hamiltonians     $H_{k}$
may be  computed from
\be
H_{k} = {1 \over k} Res \left( L^{k}\right) \, ,
\ee
in which $Res$ denotes the coefficient standing in
$\partial^{-1}$   term.

        For the subsequent discussion let us explicitly presents the
equations (6) for the two--component KP hierarchy in the two
particular cases.

For $k=2$  these equations are in the form
\be
q_{i_{t}} = q_{ixx} + 2 q_{i} \sum\limits^{m}_{i=1}
q_{s}r_{s} \, ,
\ee
\be
r_{i_{t}} = - r_{ixx} - 2 r_{i} \sum\limits^{m}_{i=1}
q_{s}r_{s}
\,  .
\ee
This is a vector generalization of the Nonlinear Schr\"{o}dinger
equation
considered first time in [49].
\\
For $k =3$
\be
q_{i_{t}} = q_{ixxx} + 3 q_{i} \sum\limits^{m}_{s=1}
q_{sx}r_{s} + 3 q_{ix} \sum\limits^{m}_{s=1}q_{s}r_{s} \, ,
\ee
\be
r_{i_{t}} = r_{ixxx} + 3 r_{i} \sum\limits^{m}_{s=1}
q_{s}r_{sx} + 3 r_{ix} \sum\limits^{m}_{s=1}q_{s}r_{s}
\,  .
\ee
These equations could be further restricted to the known soliton
equation.
Indeed, assuming that m=1 we obtain that equations (12) - (13)
  reduce to the
usual Nonlinear Schr\"{o}dinger equation while the eqs. (14)-(15)
 for $q=r $
to the modified Korteweg - de Vries equation or for $r=1$  to the
Korteweg - de Vries equation.

\section{The extended supersymmetrization of the multicomponent KP
hierarchy. }

        The basic objects in the supersymmetric analysis are the
superfield
and the supersymmetric derivative. We will deal with the so called
extended
$N=2$
 supersymmetry for which superfields are the superfermions
or the  superbosons depending,  in addition to
x  and t,
    upon two anticommuting variables,
    $\theta_{1}$     and
      $\theta_{2}$,
      $(  \theta_{2}\theta_{1} = -   \theta_{1}
\theta_{2},   \theta_{1}^{2} =   \theta_{2}^{2} = 0)$.
Their Taylor expansion with respect to the $\theta$ is
\be
\phi(x,   \theta_{1},  \theta_{2})
= w(x) + \theta_{1} \zeta_{1}(x) +
  \theta_{2}\zeta_{2}(x) +
    \theta_{2}\theta_{1}u(x),
    \ee
    where the fields
    $w, u$,    are to be interpreted as the boson (fermion) fields
for the superboson (superfermion) field
$\zeta_{1}, \zeta_{2}$,  as the fermions (bosons) for the
for the superboson (superfermion) respectively. The superderivatives
are
defined as
\be
{\cal D}_{1} =\partial_{\theta_{1}} + \theta_{1}\partial \, ,
\qquad \qquad
{\cal D}_{2} =\partial_{\theta_{2}} + \theta_{2}\partial,
\ee
with the properties
\be
{\cal D}_{2}{\cal D}_{1} +
{\cal D}_{1} {\cal D}_{2} = 0 \, ,
\qquad \qquad
{\cal D}_{1}^{2} = {\cal D}_{2}^{2} = 0.
\ee
        Below we shall use the following notation: $({\cal
D}_{i}{F})$
denotes the outcome of the action of the superderivative on the
superfield ${F}$,
while ${\cal D}_{i} {F}$ denotes the action itself of the
superderivative on the superfield ${F}$.

     The principal problem in the supersymmetrization of the soliton
equations could be formulated as follows :
if we know the evolution equation for the classical function u and its
(bi) hamiltonian structure or its Lax pair, how is its possible to
obtain
the evolution equation on the supermultiplet
$\Phi$   which contains the classical
function $u$? This problem has its own history and at the moment we
have no
an unique solution. We can distinguish three different methods of the
supersymmetrization, as for example the algebraic,
geometric and direct method.

        In the first two cases we are looking for the symmetry group
of the
given equation and then we replace this group by the corresponding
SUSY group. As a final product we are able to obtain the SUSY
generalization
of the given equation. The classification as  the algebraical or
geometrical approach is connected with the kind of symmetry which
appears
on the classical level. For example, if our classical equation could
be
described  in terms of the geometrical object then the simple
exchange
of the classical symmetry group of this object onto SUSY partner
justify
the name geometric. In the algebraic case,  we are looking for the
symmetry group of this equation without any reference to its
geometrical origin. This strategy could be applied to the so called
hidden symmetry as for example in the case of the Toda lattice [54].

        These methods have both advantages and disadvantages.
For example,
sometimes we obtain the fermionic extensions of the given equations
only
[44,54]. In the case of the
extended supersymmetric Korteweg-de Vries equation  we have three
different
fully SUSY extensions,  however only one of them fits to these two
classifications [26--28,31].

        It seems that the most difficult problem in these approaches
is the
explanation why a'priori SUSY extension of the classical system of
equation
should be connected with the SUSY extension of the classical symmetry
of
these equations. By these reasons we prefer to use the direct
approach in which
we simply replace all objects which appear in the evolution equation
by all possible supermultiplets and superderivatives in such a way
that
to conserve the gradations of the equation. This is highly non unique
procedure and we obtain a lot
of different possibilities. However this arbitrariness can be
restricted
if we additionally investigate its super-bi-hamiltonian structure or
try to
find its supersymmetric Lax pair. In many cases this manner bring the
success [34,35,40--48]. In the next we utilize this way.

        Let us now start our considerations of trying to find the Lax
operator
for the multicomponent SUSY KP hierarchy. The direct method suggests
to assume that L depends on the vectors supermultiplets $F$, $G$
 its supersymmetric
derivatives and on the derivative and superderivatives in such a way
that
finally it has the gradation 1.
Therefore we postulate that the Lax pair is an operator in the form

\be
\begin{array}{l}
L =  L\left(\partial, D_{1}, D_{2}, {F}, {G} \right).
\end{array}
\ee
In order to specify this form we have to assume the gradations of the
supermultiplets ${F}$ and
${G}$. However we quickly recognize that we encounter
three different possibilities of the gradations of ${F}$,
${G}$:
\begin{description}
\item{}1.) All  ${F}$,
${G}$ are superfermions with the gradation 1/2,
\item{}
2.) All ${F}$,
 ${G}$ are superbosons with the following gradation:
   ${F}$ has 0 while
   ${G}$ has 1 (or symmetrically).
\item{}
3.) The mixture of both previous possibilities in other words some of
the
    ${F}$ and
    ${G}$ are superbosons and the rest are superfermions.

        In the next sections we investigate in more details these
possibilities.
\end{description}

\section{The superfermionic approach.}

        We now assume that the components of the vectors
supermultiplets
$F$ and $G$ are superfermions which could be written down as
\be
F_{i} = \zeta^{1}_{i} +
\theta_{1} f^{1}_{i}
+ \theta_{2} f^{2}_{i}
+ \theta_{2}\theta_{1} \zeta^{2}_{i},
\ee
\be
G_{i} = \eta^{1}_{i} +
\theta_{1} g^{1}_{i}
+ \theta_{2} g^{2}_{i}
+ \theta_{2}\theta_{1} \eta^{2}_{i},
\ee
where $f^{i}_{j}, g^{k}_{i}$  are usual classical functions while
$\zeta^{k}_{j},\eta^{k}_{i}$
are Grassmannian valued functions. The Lax operator we choose in such
a way
that contains all possible combinations of "variables" in the (20) in
such manner that each term has gradation 1. Then using the symbolic
language REDUCE we verified that the following operator
\be
L= \partial + \sum\limits^{k}_{i=1} F_{i} \cdot
\partial^{-1} \cdot D_{1}\cdot D_{2} \cdot G_{i},
\ee
generate extended supersymmetric multicomponent KP hierarchy.
Indeed,
its second flow is
\be
F_{i_{t}} = F_{ixx} +
2 \sum\limits^{k}_{s=1}
F_{s}\left(
{\cal D}_{1} {\cal D}_{2} {G}_{s} {F}_{i}\right)
- F_{i}
\left( \sum\limits^{k}_{s=1} F_{s} G_{s} \right)^{2}
\, ,
\ee
\be
G_{i_{t}} = G_{ixx} +
2 \sum\limits^{k}_{s=1}
G_{s}\left(
{\cal D}_{1} {\cal D}_{2} {G}_{i} {F}_{s}\right)
+ G_{i}
\left( \sum\limits^{k}_{s=1} F_{s} G_{s} \right)^{2}\, ,
\ee
while the third is
\be
\begin{array}{ll}
F_{i_{t}} = & F_{ixxx} +
3 \sum\limits^{k}_{j=1}
\bigg\{
\left(
{\cal D}_{1} {\cal D}_{2} {G}_{j} {F}_{ix}\right) +
\left({\cal D}_{1} {\cal D}_{2} {G}_{j} {F}_{i}\right)
F_{jx} -
\cr\cr
&
\sum\limits^{k}_{l=1}
\left[ \left(
{\cal D}_{1} {G}_{j} {F}_{l}\right)
\left({\cal D}_{1} {G}_{l} {F}_{i}\right)
F_{j} +
 \left({\cal D}_{2} {G}_{j} {F}_{l}\right)
 \left({\cal D}_{2} {G}_{l} {F}_{i}\right)
 F_{j} \right]
 \bigg\} - 3F_{ix} Z \, ,
 \end{array}
 \ee
 \be
 \begin{array}{ll}
G_{i_{t}} = & G_{ix} +
3 \sum\limits^{k}_{j=1}
\bigg\{
\left(
{\cal D}_{1} {\cal D}_{2} {G}_{ix} {F}_{j}\right) +
G_{j}\left({\cal D}_{1} {\cal D}_{2} {G}_{i} {F}_{j}\right)
G_{jx} +
\cr\cr
&
+ \sum\limits^{k}_{l=1}
 \left(
{\cal D}_{1} {G}_{i} {F}_{j}\right)
\left({\cal D}_{1} {G}_{i} {F}_{l}\right)
G_{l} +
 \left({\cal D}_{2} {G}_{i} \right)
 \left({\cal D}_{2} {G}_{j} {F}_{l}\right)
 G_{l}  \bigg\} - 3G_{ix} Z\, ,
 \end{array}
 \ee
where
\be
Z = \sum\limits^{k}_{i,j=1} F_{i} G_{i} F_{j}G_{j}.
\ee
Let us now discuss several particular cases of the equations (24--27).
For k=1,  equations (24--25) reduces to
\be
F_{t} = F_{xx} + 2 F\left(
 {\cal D}_{1} {\cal D}_{2}{G} F \right)\, ,
\ee
\be
G_{t}= -
G_{xx} - 2 G \left({\cal D}_{1} {\cal D}_{2}{G} F \right)
\, ,
\ee
In the components,  using (21--22), we obtained  that eqs. (29--30)
are equivalent with

\be
\zeta^{1}_{t} = \zeta^{1}_{xx} + 2 \zeta^{1}
\left( \eta^{1}\zeta^{2} + f^{1}g^{2} - f^{2}g^{1} \right),
\ee
\be
f^{1}_{t} = f^{1}_{xx} - 2 \zeta^{1}
\left( g^{2}\zeta^{1} - f^{2}\eta^{2} \right)_{x}
+ 2 f^{1}
\left(\eta^{1}\zeta^{2} + \eta^{2} \zeta^{1} + f^{1}g^{2} - f^{2}g^{1}
\right),
\ee
\be
f^{2}_{t} = f^{2}_{xx} + 2 \zeta^{1}
\left( g^{1}\zeta^{1} - f^{1}\eta^{1} \right)_{x}
+ 2f^{2}
\left(\eta^{1}\zeta^{2} + \eta^{2} \zeta^{1} + f^{1}g^{2} - f^{2}g^{1}
\right),
\ee
\be
\begin{array}{ll}
\zeta^{2}_{t} = &  \zeta^{2}_{xx} - 2 \zeta^{1}
\left( \eta^{1}\zeta^{1}\right)_{xx}
+ 2 f^{1}
\left(g^{1}\zeta^{1} - f^{1}\eta^{1} \right)_{x} +
\cr\cr
&  2 f^{2}
\left( g^{2}\zeta^{1} - f^{2}\eta^{1}\right)_{x}
+ 2 \zeta^{2}
\left( \eta^{2}\zeta^{1} + f^{1}g^{2} - f^{2}g^{1}
\right),
\end{array}
\ee

\be
\eta^{1}_{t} = - \eta^{1}_{xx} - 2 \eta^{1}
\left( \eta^{2}\zeta^{1} + f^{1}g^{2} - f^{2}g^{1} \right),
\ee
\be
g^{1}_{t} = - g^{1}_{xx} + 2 \eta^{1}
\left( g^{2}\zeta^{1} - f^{2}\eta^{1} \right)_{x}
- 2 g^{1}
\left(\eta^{1}\zeta^{2} + \eta^{2} \zeta^{1}
+ f^{1}g^{2} - f^{2}g^{1}
\right),
\ee
\be
g^{2}_{t} = - g^{2}_{xx} - 2 \eta^{1}
\left( g^{1}\zeta^{1} - f^{1}\eta^{1} \right)_{x}
- 2 g^{2}
\left(\eta^{1}\zeta^{2} + \eta^{2} \zeta^{1}
+ f^{1}g^{2} - f^{2}g^{1}
\right),
\ee
\be
\begin{array}{ll}
\eta^{2}_{t} = & - \eta^{2}_{xx} + 2 \eta^{1}
\left( \eta^{1}\zeta^{1}\right)_{xx}
- 2 g^{1}
\left(g^{1}\zeta^{1} - f^{1}\eta^{1} \right)_{x} -
\cr\cr
&  2 g^{2}
\left( g^{2}\zeta^{1} - f^{2}\eta^{1}\right)_{x}
+ 2 \eta^{2}
\left( \eta^{1}\zeta^{2} + f^{1}g^{2} - f^{2}g^{1}
\right).
\end{array}
\ee

As we see,  this system of equations  can be interpreted as the
extended
supersymmetric Nonlinear Schr\"{o}dinger equation which have been
widely
discussed recently [36-41,48]. The bosonic part (in which all fermions
fields vanishes) give us the equations (7) for m=2 with the following
identifications
\be
f^{1} = g_{1} \, , \quad
f^{2} = g_{2} \, , \quad
q^{1} = - r_{2} \, , \quad
q^{2} = r_{1}.
\ee
Interestingly, our Lax operator in the bosonic limit for k=1 does not
reduce to the scalar Lax pair (4). In our case,  it has a matrix form
\be
L = \pmatrix{\partial + q_{1} \partial^{-1} r_{1}\, ,
& q_{1}\partial^{-1} r_{2} \cr\cr
 q_{2}\partial^{-1} r_{1}\, ,
 & \partial + q_{2} \partial^{-1} r_{2}}.
\ee
In this way,  we have shown that our one--component extended
supersymmetric
KP hierarchy in the bosonic sector is equivalent with the usual
two--component KP hierarchy. Moreover,  in this bosonic sector, our
equations
constitute the bi-Hamiltonian structure given by (6-11),  but we are
not
able to find its supersymmetric bihamiltonian counterparts. This fact
probably is connected with the nonexistence of the extended (N=2)
supersymmetric SL(2,C) Kac-Moody algebra [30]. This SL(2,C) Kac-Moody
algebra plays
a crucial meaning in the AKNS approaches and to its bi-Hamiltonian
structure
as it was shown in the previous section.
Moreover,  the applications of the direct method to the
supersymmetrization
of the formula (11) also does not give us the correct solution, what
we
have checked using the symbolic computation program REDUCE.

        On the other hand our equations are Hamiltonian equations
which can be written as
\be
\pmatrix{F \cr\cr G }_{t_{k}} =
\pmatrix{0, & I \cr\cr
- I, & 0 }
\displaystyle\pmatrix{ \displaystyle{\delta H_{k} \over \delta F}
\cr\cr
\displaystyle{\delta H_{k} \over \delta G} },
\ee
where
$F=(F_{1},F_{2},...F_{k})^{t}$,
$G=(G_{1},G_{2},...G_{k})^{t}$  and  $I$  is a  $k \times k$
  identity
matrix. The Hamiltonians $H_{k}$ can be computed by using the formula
(12)
in which  now the$Res$  denotes the coefficient standing in
$\partial^{-1}D_{1}D_{2}$       term.

        Let us now discuss the equations (25-26) for k=1.  In this
case
they  reduce to
\be
F_{t} = F_{xxx} + 3 \left[ \left(
 {\cal D}_{1} {\cal D}_{2}{G} {F}
 _{x} \right) F
 +
 \left({\cal D}_{1} {\cal D}_{2}{G} {F}
 _{x} \right) F_{x} \right],
\ee
\be
G_{t} = G_{xxx} + 3 \left[ \left(
 {\cal D}_{1} {\cal D}_{2}{G}_{x} {F}
  \right) G
 +
 \left({\cal D}_{1} {\cal D}_{2}{G} {F}
 _{x} \right) G_{x} \right],
\ee
with the following bosonic sector
\be
f_{1t} = f_{1xxx} - 3g_{1}\left(f_{2}f_{1}\right)_{x}
+ 3 g_{2}\left(f^{2}_{1}\right)_{x},
\ee
\be
f_{2t} = f_{2xxx} + 3g_{2}\left(f_{1}f_{2}\right)_{x}
- 3 g_{1}\left(f^{2}_{2}\right)_{x},
\ee
\be
g_{1t} = g_{1xxx} + 3f_{1}\left(g_{1}g_{2}\right)_{x}
- 3 f_{2}\left(g^{2}_{1}\right)_{x},
\ee
\be
g_{2t} = g_{2xxx} - 3f_{2}\left(g_{1}g_{2}\right)_{x}
+ 3 f_{1}\left(g^{2}_{2}\right)_{x}.
\ee

        This system of equation can be considered as the vector
generalizations of the Modified Korteweg - de Vries equation. Now we
can
investigate different reductions of the eqs. (44-47) to much simpler
equations. For example, by assuming that
\be
g_{1} = f_{1} = f_{2}, \qquad\qquad  g_{2} =0,
\ee
we obtain the usual Modified Korteweg - de Vries equation.

        To finish this section let us notice that this superfermionic
manner discussed in this section allows us to obtain some extension
of
the usual system of equations by incorporating anticommuting
functions
but we do not change the usual multicomponent K-P hierarchy. We show
in
the next sections that superbosonic or mixed ways of
supersymmetrizations generalize our usual multicomponent K-P hierarchy
in the class of the usual commuting functions.

\section{The superbosonic approach.}

We now assume that the components of the vector supermultiplet
$F$ and $G$ are superbosons and could be expressed as
\be
F_{i} = f^{1}_{i} + \theta_{1}\zeta^{1}_{i} +
\theta_{2}\zeta^{2}_{i} + \theta_{2}\theta_{1}f^{2}_{i},
\ee
\be
G_{i} = g^{1}_{i} + \theta_{1}\eta^{1}_{i} +
\theta_{2}\eta^{2}_{i} + \theta_{2}\theta_{1}g^{2}_{i},
\ee
where $\zeta^{j}_{i}$    and
$\eta^{j}_{i}$    are Grassmann valued functions while
$f^{j}_{i}$, $g^{j}_{i}$        are
usual commuting functions. In order to find the proper Lax operator in
this case we assume the following gradation  on the functions
\be
\begin{array}{l}
\deg \left( f^{1}_{i} \right) = 0 \, ,
\qquad
\deg \left( \zeta^{j}_{i} \right) = 0.5 \, ,
\qquad
\deg \left( f^{2}_{i} \right) = 1,
\cr\cr
\deg \left( g^{1}_{i} \right) = 1 \, ,
\qquad
\deg \left( \eta^{j}_{i} \right) = 1.5 \, ,
\qquad
\deg \left( g^{2}_{i} \right) = 2, .
\end{array}
\ee
Notice that it is possible also to assume the symmetrical gradation
in which we replace
$f \to g$,
$\zeta \to \eta$
             but we will not consider such possibility
because we obtain the same information as in the considered  case.

        We postulate  the Lax operator exactly as in the formula (22)
and
interestingly in this case we obtain the same flows,
      where now in contrast
to eq. (23)
$ F$
 and  $G$
  are superbosons. Therefore, they have different expansions
in the components. Let us consider more carefully two particular
cases
($ k=1 $) of these flows.
The second flow is
\be
{d \over dt} F = F_{xx} - G^{2} F^{3} + 2 F \left(
{\cal D}_{1} {\cal D}_{2} G F \right),
\ee
\be
{d \over dt} G = G_{xx} - G^{3} F^{2} - 2 G \left(
{\cal D}_{1} {\cal D}_{2} G F \right).
\ee
It is the extended supersymmetric Nonlinear Schr\"{o}dinger Equation
considered
in [48].
The third flow is
\be
{d \over dt} F = F_{xxx}+ 3 F_{x}
\left(
{\cal D}_{1} {\cal D}_{2} G F \right)
+ 3 F
\left(
{\cal D}_{1} {\cal D}_{2} G F_{x} \right)
-3F^{2}G^{2} F_{x},
\ee
\be
{d \over dt} G = G_{xxx}+ 3 G_{x}
\left(
{\cal D}_{1} {\cal D}_{2} G F \right)
+ 3 G
\left(
{\cal D}_{1} {\cal D}_{2} G_{x} F \right)
-3F^{2}G^{2} G_{x}.
\ee
{}From the last equation it follows that for   $F=-1 $, our
equations reduces to the supersymmetric Korteweg - de Vries.
As it is known there are three different generalization of the
extended supersymmetric KdV equation which have the Lax representation
[27--29,42,46] and this can be compactly written down   as
\be
{d \over dt} G = \left(
- G_{xx} +
3 G \left( {\cal D}_{1} {\cal D}_{2} {G} \right) +
{1\over 2} (\alpha - 1)
\left( {\cal D}_{1}{\cal D}_{2} {G}^{2} \right)
+ \alpha {G}^{3} \right)_{x}
\ee
Here, $ \alpha $ is just a free parameter which enumerates these
three different
cases. Our case corresponds to  $ \alpha=1 $, after rescaling the
time and
transforming ${G}$ into ${- G}$.
In the paper [46] the present author considered the nonstandard Lax
representations for this equation. Here, as the byproduct of our
analysis
we obtained
the usual Lax representation for this equation which could be
connected
with the extended supersymmetric AKNS approach. Indeed,  our Lax
operator
in this case takes the form
\be
L=\partial - \partial^{-1} {\cal D}_{1} {\cal D}_{2} G\,
\ee
with the following flow
\be
L_{t} = \left[\left(L^{3}\right)_{+}, L\right]\, .
\ee
Unfortunately,  similarly to the superfermionic case considered
in the previous chapter we have not
found the bihamiltonian structure of this equation.

\section{The superfermionic and superbosonic approach.}

        We are now able to consider the mixed approaches to the
construction of the SUSY multicomponent KP hierarchy. Therefore we
consider now the following SUSY lax operator
\be
L = \partial + \sum\limits^{k}_{i=1}
F_{i} \partial^{-1} D_{1}D_{2}G_{i}
+ \sum\limits^{m}_{j=1} B_{i} \partial^{-1}
D_{1} D_{2} C_{i},
\ee
where now F and G  are vector superfermions with the expansions (20-
21)
while B and C are superbosons
with  the following expansions (47-48).
        Using the same technique as in the previous sections we
computed the
second and third flows but the final formulas are complicated. Hence
we present the second flow only which can be written down as
\be
{d \over dt} F_{i} = F_{ixx} - F_{i} Z + 2
\sum\limits^{k}_{l=1}
F_{l} \left({\cal D}_{1} {\cal D}_{2} G_{l} F_{i} \right) + 2
\sum\limits^{m}_{j=1}
B_{j} \left({\cal D}_{1} {\cal D}_{2} C_{j} F_{i} \right),
\ee
\be
{d \over dt} B_{j} = B_{jxx} - B_{j} Z + 2
\sum\limits^{k}_{l=1}
F_{l} \left({\cal D}_{1} {\cal D}_{2} G_{l} B_{j} \right) + 2
\sum\limits^{m}_{s=1}
B_{s} \left({\cal D}_{1} {\cal D}_{2} C_{s} B_{j} \right),
\ee

\be
{d \over dt} G_{i} = - G_{ixx} + G_{i} Z - 2
\sum\limits^{k}_{l=1}
G_{l} \left({\cal D}_{1} {\cal D}_{2} G_{i} F_{l} \right) - 2
\sum\limits^{m}_{j=1}
C_{j} \left({\cal D}_{1} {\cal D}_{2} G_{i} B_{j} \right),
\ee
\be
{d \over dt} C_{j} = - C_{jxx} + C_{j} Z - 2
\sum\limits^{k}_{l=1}
G_{l} \left({\cal D}_{1} {\cal D}_{2} C_{j} F_{l} \right) - 2
\sum\limits^{m}_{s=1}
C_{s} \left({\cal D}_{1} {\cal D}_{2} C_{j} B_{s} \right) .
\ee
where

\be
{Z} = \left( \sum\limits^{k}_{l=1} F_{i}G_{i}
+ \sum\limits^{m}_{j=1}
B_{j} C_{j} \right)^{2}
\ee

        As we see, the last system of equations describes  the huge
class  of  interacting fields.  In some sense, it describes the
interaction of the superfermions with the superbosons.

\section{Concluding remarks. }

        We have constructed the extended
supersymmetric version of the multicomponent K-P hierarchy in three
different ways.  We obtained a new class of integrable equation
for which we were able to constructed the Lax operator and showed
that they are
Hamiltonian equations.  Moreover, due to the existence of the
Lax operator,
we obtained the infinite number of conserved currents for our
generalizations.
Unfortunately we could not prove that these currents are in
involution.

        In the soliton theory, in order to prove the involution of the
conserved currents, we utilize the recursion operator.  Magri [53]
has shown that
such recursion operator could be constructed if we know the
bihamiltonian
structure.  However in our case we could not find such bihamiltonian
structure.
It does not mean that our system does not possess the recursion
operator.
The excellent example of the situation, where we do not known the
bihamiltonian
structure, but we  known the recursion operator is the Burgers
equation [55].
Therefore, it seems reasonable that if we wish to find the recursion
operator
for our supersymmetric generalizations we should try, to follow the
Burgers approach.


\begin{thebibliography}{99}
\bibitem{1. }  L. Faddeev and L. Takhtajan: ``Hamiltonian
Methods in the Theory of
        Solitons ",  Springer (1987); A. Das:  ``Integrable Models",
World
        Sci. (1989); M. Ablowitz and H. Segur: ``Solitons and the
Inverse
        Scattering Transform",  SIAM Philadelphia (1981).

\bibitem{2. }  A. Polyakov: in ``Fields,  Strings and critical
Phenomena",
 ed. E. Brezin and
        J. Zinn-Justin,  North Holland (1989).

\bibitem{3. }  D. Gross and A. Migdal Nucl. Phys.  {\bf B340}, 333
(1990);
E. Brezin and V. Kazakov
        Phys. Lett., {\bf  236B},  144 (1990).

\bibitem{4. }  S. Manakov,  S. Novikov,  L. Pitaevski and V.
Zakharov:
``Soliton Theory: The
        Inverse Problem",  Nauka,   Moscow (1980).

\bibitem{5. }  L.A. Dickey: ``Soliton Equations and
Hamiltonian Systems",  World
        Scientific, Singapore (1991).

\bibitem{6. }  E. Date,  M. Jimbo,  M. Kashiwara and T. Miwa,
in: ``Nonlinear Integrable
        Systems - Classical and Quantum Theory",  ed.  by M. Jimbo and
        T. Miwa, World Scientific,  Singapore 1983,  p. 39.

\bibitem{7. }  M. Jimbo and T. Miwa,  Publ.  RIMS,  Kyoto Univ.
{\bf 19}, 943 (1983).

\bibitem{8. }  Y. Ohta,  J. Satsuma,  D. Takahashi,  and T. Tokihiro,

Prog. Theor. Phys. Suppl.
      {\bf  94}, 219 (1988).

\bibitem{9. }  B.G. Konopelchenko and W. Strampp,  Inverse
Problems {\bf 7},
  L17 (1991).

\bibitem{10. }  B.G. Konopelchenko,  J. Sidorenko  and W. Strampp,
Phys. Lett.,
 {\bf A 157},  17 (1991).

\bibitem{11. }  J. Sidorenko and W. Strampp,  Inverse Problems
{\bf 7},
L37 (1991).

\bibitem{12. }  Y. Cheng and Y. Li,  Phys. Lett.,  {\bf A157}, 22
(1991).

\bibitem{13. } Y. Cheng,  J. Math. Phys., {\bf 33}, 3774 (1992).

\bibitem{14. }  J. Sidorenko and W. Strampp,  J. Math. Phys.
{\bf 34}, 1429 (1993).

\bibitem{15. }  W. Oevel and W. Strampp,  Commun. Math. Phys.
{\bf 157}, 51 (1993).

\bibitem{16. }  N. Yajima and M. Oikawa,  Progr. Theor. Phys.
{\bf 56},  1719 (1976).

\bibitem{17. }  J. Wess and J. Bagger:  ``Supersymmetry and
Supergravity",
Princeton, NJ (1982);

\bibitem{18. }  S. Ferrara and J.G. Taylor: ``Introduction to
Supergravity"
  (in russian)
         Moscow 1985.

\bibitem{19. }  B. Kupershmidt: ``Elements of Superintegrable
Systems",
Kluwer 1987.

\bibitem{20. }  M. Chaichian,  P. Kulish, Phys. Lett. {\bf 18B}, 413,
(1978).

\bibitem{21. }  R. D'Auria and S. Sciuto,  Nucl.  Phys.
{\bf B171}, 189, (1980).

\bibitem{22. } M. Gurses and O. Oguz,  Phys. Lett. {\bf 108A}, 437,
(1985).

\bibitem{23. } Y. Manin and R. Radul,  Commun. Math. Phys.  {\bf
98}, 65 (1985).

\bibitem{24. } C. Morosi and L. Pizzochero,  Commun. Math. Phys.
{\bf 158}, 267 (1993).

\bibitem{25. } C. Morosi and L. Pizzochero,  J. Math. Phys.,
{\bf 35}, 2397, (1994).

\bibitem{26. } C. Morosi and L. Pizzochero: ``A Fully Supersymmetric
AKNS Theory ",
        preprint of Dipartimento di Matematica, Politecnico di
Milano.
        April 1994,  to appeare in Commun. Math. Phys. (1995).

\bibitem{27. } P. Mathieu,  J. Math. Phys.,
{\bf 29}, 2499, (1988).

\bibitem{28. } C.A. Laberge and P. Mathieu,  Phys. Lett.
{\bf 215B}, 718, (1988).

\bibitem{29. } P. Labelle and P. Mathieu,  J. Math. Phys.,
{\bf 32}, 923, (1991).

\bibitem{30. } M. Chaichian and J. Lukierski,  Phys. Lett.,
{\bf 183B}, 169, (1987).

\bibitem{31. } T. Inami and H. Kanno,  Commun. Math. Phys.
{\bf 136}, 519 (1991).

\bibitem{32. } S.K. Nam,  Intern. J. Mod. Phys., {\bf 4}, 4083,
(1989).

\bibitem{33. } K. Hiutu and D. Nemeschansky,  Mod. Phys. Lett.
 {\bf A6},  3179,  (1991).

\bibitem{34. } C.M. Yung,   Phys. Lett.  {\bf 309B}, 75, (1993).

\bibitem{35. } E. Ivanov and S. Krivons,  Phys. Lett.
{\bf 291B}, 63, (1992).

\bibitem{36. } P. Kulish, Lett. Math. Phys. {\bf 10}, 87, (1985).

\bibitem{37. } G.H.M. Roelofs and P.H.M.  Kersten, Journ.  Math.
Phys.
{\bf 33}, 2185 (1992).

\bibitem{38. } J.C. Brunelli and A. Das,  Journ. Math. Phys.
{\bf 36}, 268, (1995).

\bibitem{39. } F. Toppan,  Int. Journ. Mod. Phys.
{\bf A10}, 895, (1995).

\bibitem{40. } S. Krivonos and A. Sorin: ``The minimal N=2
superextension of the NLS
        equation",  hep-th/9504084 to appeare in Phys. lett. B.

\bibitem{41. } S. Krivonos,  A. Sorin and F. Toppan: ``On the Super-
NLS
Equation and its
        Relation with N=2 Super-KdV within Coset Approach",  hep-
th/9504138.

\bibitem{42. } W. Oevel and Z. Popowicz,  Commun. Math. Phys.
{\bf 139}, 441, (1991).

\bibitem{43. } Z. Popowicz,  J. Phys. A:Math. Gen.
{\bf 19}, 1495, (1986).

\bibitem{44. } Z. Popowicz,  J. Math. A:Math. Gen.
{\bf 23}, 1127, (1990).

\bibitem{45. } Z. Popowicz,  Phys. Lett.
{\bf 319B}, 478,  (1993).

\bibitem{46. } Z. Popowicz,  Phys. Lett.  {\bf 174A}, 411, (1993).

\bibitem{47. } Z. Popowicz,  Int. Jour. Mod. Phys.
{\bf 9}, 2001, (1994).

\bibitem{48. } Z. Popowicz,  Phys. Lett.
{\bf 194A}, 375, (1994).

\bibitem{49. } S. Manakov, Sov. Phys. JEPT
{\bf 38}, 248 (1974).

\bibitem{50. } V.K. Melnikv,  Commun. Math. Phys.
{\bf 112}, 639, (1987).

\bibitem{51. } Q.P. Liu: ``Hamiltonian Structure of Multicomponent
Constrained KP
        Hierarchy",  hep-th/9502076.

\bibitem{52. } H. Aratyn,  J.F. Gomes and A.H. Zimerman:
``Affine Lie Algeraic Origin
        of Constrained KP Hierarchies", hep-th/9408104 to appeare in
        Journ. of Math. Phys.
\bibitem{53} F. Magri,  J. Math. Phys.
{\bf 19}, 1156 (1978).

\bibitem{54} M. Olshanetsky,  Commun. Math. Phys.
{\bf 30} (1975) 1975.

\bibitem{55} W. Oevel and W. Strampp,  Commun. Math. Phys.
{\bf 157},  51 (1983).
\end{thebibliography}
\end{document}